\documentclass[twocolumn,amsmath,aps,fleqn]{revtex4}
\usepackage{graphicx,amssymb}
\begin{document}
\newcommand{\be}{\begin{equation}}
\newcommand{\ee}{\end{equation}}
\newcommand{\bq}{\begin{eqnarray}}
\newcommand{\eq}{\end{eqnarray}}
\newcommand{\bsq}{\begin{subequations}}
\newcommand{\esq}{\end{subequations}}
\newcommand{\bc}{\begin{center}}
\newcommand{\ec}{\end{center}}
\newcommand {\R}{{\mathcal R}}
\newcommand{\al}{\alpha}
\newcommand\lsim{\mathrel{\rlap{\lower4pt\hbox{\hskip1pt$\sim$}}
    \raise1pt\hbox{$<$}}}
\newcommand\gsim{\mathrel{\rlap{\lower4pt\hbox{\hskip1pt$\sim$}}
    \raise1pt\hbox{$>$}}}

\title{Interacting dark energy: the role of microscopic feedback in the dark sector}

\author{P.P. Avelino}
\email[Electronic address: ]{pedro.avelino@astro.up.pt}
\affiliation{Instituto de Astrof\'{\i}sica e Ci\^encias do Espa{\c c}o, Universidade do Porto, CAUP, Rua das Estrelas, PT4150-762 Porto, Portugal}
\affiliation{Centro de Astrof\'{\i}sica da Universidade do Porto, Rua das Estrelas, PT4150-762 Porto, Portugal}
\affiliation{Departamento de F\'{\i}sica e Astronomia, Faculdade de Ci\^encias, Universidade do Porto, Rua do Campo Alegre 687, PT4169-007 Porto, Portugal}

\date{\today}
\begin{abstract}

We investigate the impact on the classical dynamics of dark matter particles and dark energy of a non-minimal coupling in the dark sector, assuming that the mass of the dark matter particles is coupled to a dark energy scalar field. We show that standard results can only be recovered if the space-time variation of the dark energy scalar field is sufficiently smooth on the characteristic length scale of the dark matter particles, and we determine the associated constraint dependent on both the mass and radius of the dark matter particles and the coupling to the dark energy scalar field. We further show, using field theory numerical simulations, that a violation of such constraint results in a microscopic feedback effect strongly affecting the dynamics of dark matter particles, with a potential impact on structure formation and on the space-time evolution of the dark energy equation of state.

\end{abstract}
\maketitle

\section{\label{intr}Introduction}

Cosmological observations provide overwhelming evidence that our Universe is currently undergoing an inflationary phase \cite{Suzuki:2011hu,Anderson:2012sa,Parkinson:2012vd,Hinshaw:2012aka,Ade:2013zuv,Planck:2015xua}. In the context of Einstein's General Relativity, this accelerated expansion can only be explained if the Universe is presently dominated by an exotic Dark Energy (DE) form, violating the strong energy condition \cite{Copeland:2006wr,Frieman:2008sn,Caldwell:2009ix,Li:2011sd,Bamba:2012cp}.  While DE explains the observed dynamics of the Universe on cosmological scales, a non-relativistic Dark Matter (DM) component is required in order to account for the observed dynamics of cosmological perturbations over a wide range of scales. Together DM and DE seem account for about 95$\%$ of the total energy density of the Universe at the present time. However, our knowledge of DM and DE is indirect, relying only on their contribution to the gravitational field. Unveiling the nature of DM and DE is therefore one of the most ambitious challenges of fundamental physics. 

The possibility of a non-minimal coupling in the dark sector \cite{Wetterich:1994bg,Amendola:1999er,Zimdahl:2001ar,Farrar:2003uw} is an exciting topic of current research. Such a coupling could affect both the background evolution of the Universe as well as on the growth of cosmological perturbations \cite{Gumjudpai:2005ry,Pettorino:2008ez,CalderaCabral:2008bx,CalderaCabral:2009ja,Baldi:2010pq}, with a potential impact on the redshift dependence of the apparent magnitude of type Ia supernovae, the Cosmic Microwave Background (CMB) anisotropies, baryonic acoustic oscillations, redshift-space distortions and weak gravitational lensing (see, e.g., \cite{Ade:2015yua} and references therein).  A non-minimal coupling between DM and DE may naturally lead to a fractional matter abundance at the present time significantly different from the one obtained using the information contained in the CMB temperature power spectrum assuming no interaction between DM and DE \cite{Avelino:2012tc}, which can help resolving the tension between CMB  and local constraints on the value of the Hubble parameter. In fact preliminary indications of a late-time non-minimal interaction between DM and DE have been recently reported in  \cite{Salvatelli:2013wra,Pettorino:2013oxa,Salvatelli:2014zta,Abdalla:2014cla} (see also \cite{Ade:2015yua}). The coupling of DE with other fields has also been investigated in the context of varying fundamental constants (see, e.g., \cite{Avelino:2006gc}) and growing neutrino models (see, e.g., \cite{Wetterich:2007kr,Ayaita:2011ay}).

In previous studies, the coupling between DM and DE energy has been considered at a macroscopic level. These studies (see, e.g., \cite{Ayaita:2011ay,Faraoni:2014vra}) implicitly assume the space-time variation of the DE field to be sufficiently smooth on the characteristic length scale of the DM particles, thus neglecting any non-linear feedback at the microscopic level. In this paper we intend to bridge this gap by studying the impact of a non-minimal coupling in the dark sector on the microscopic dynamics of DM particles and DE. In Sec. \ref{sec2} we start by presenting a simple field theory model, where the mass and size of the DM particles is assumed to be a function of a DE scalar field, and we use it to derive the standard results for the dynamics of DM particles non-minimally coupled to the DE field assuming a negligible microscopic feedback. In Sec. \ref{sec3} we perform field theory numerical simulations in $1+1$ dimensions and quantify the effect of the microscopic feedback neglected in the previous section. We discuss the potential cosmological implications of our results and present the main conclusions of this work in Sec. \ref{conc}.

Throughout this paper we use units such that $c=1$, where $c$ is the value of the speed of light in vacuum, and we adopt the metric signature $(-,+,+,+)$.

\section{Interacting dark energy model\label{sec2}}

In this paper we study the microscopic dynamics of non-minimally interacting DM and DE. Although the main results derived in the paper are generic, for simplicity we shall assume DM and DE to be described by two coupled scalar fields, $\phi$ and $\varphi$, living in a $1+1$ dimensional space-time. We shall consider a class of models described by the action
\be\label{eq:L}
S=\int d^2x \, \sqrt{-g}  \, {\mathcal L}\, ,
\ee
where the Lagrangian $\mathcal L$ is given by
\be
{\mathcal L}= -\frac12 \partial_\mu \phi \partial^\mu \phi  -\frac12 \partial_\nu \varphi \partial^\nu \varphi - V(\phi,\varphi)\,. 
\ee
Here, $g=\det (g_{\mu\nu})$ and $g_{\mu\nu}$ are the components of the metric tensor.
For concreteness, we shall assume the following form of the potential
\bq
V(\phi,\varphi)&=&U(\phi,\varphi) + W(\varphi)\,,\\
U(\phi,\varphi)&=&\frac{\lambda(\varphi)}{4} (\phi^2 - \eta^2)^2\,,
\eq
where the scalar fields $\phi$ and $\varphi$ play essentially a DM and DE role, respectively, with the non-minimal interaction between them being realized through the $\lambda \equiv \lambda(\varphi)$ term. 

The equations of motion for the scalar fields 
\bq
\Box \phi&=&\frac{\partial U}{\partial\phi}\,,\\
\Box \varphi&=& \frac{d W}{d\varphi}-2\beta U\,,
\eq
may be obtained by minimizing the action with respect to variations of $\phi$ and $\varphi$. Here, $\Box \equiv \nabla_\mu \nabla^\mu$,  $\nabla_\mu$ represent the covariant derivates and
\be
\beta\equiv-\frac12\frac{d \ln \lambda}{d\varphi}\,.
\ee

The energy-momentum tensor 
\be
T^{\mu\nu}=\frac{2}{{\sqrt {-g}}} \frac{\delta({\mathcal L}{\sqrt {-g}})}{\delta g_{\mu \nu}}=2\frac{\delta{\mathcal L}}{\delta g_{\mu \nu}}+g^{\mu\nu} {\mathcal L}\,,
\ee
is given by
\be
T^{\mu\nu}=T^{\mu\nu}_{DM}+T^{\mu\nu}_{DE}\,,
\ee
with
\bq
T^{\mu\nu}_{DM}&=&\partial^\mu \phi \partial^\nu \phi -\frac{g^{\mu\nu}}{2} \left(\partial_\alpha \phi \partial^\alpha \phi+2U(\phi,\varphi)\right)\,,\label{TDM}\\
T^{\mu\nu}_{DE}&=&\partial^\mu \varphi \partial^\nu \varphi -\frac{g^{\mu\nu}}{2} \left(\partial_\alpha \varphi \partial^\alpha \varphi+2W(\varphi)\right)\,.
\eq
The energy-momentum tensors of the DM and DE components are not individually conserved since
\bq
\nabla_\nu T^{\mu\nu}_{DM}&=&Q^\mu\,, \label{DMQ}\\
\nabla_\nu T^{\mu\nu}_{DE}&=&-Q^\mu\,,
\eq
with
\be
Q^\mu=2\beta U(\phi,\varphi) \partial^\mu \varphi\,.
\ee
However, the total energy-momentum tensor is covariantly conserved ($\nabla_\nu T^{\mu\nu}=0$). Further ahead, we shall demonstrate that Eq. (\ref{DMQ}), ignoring microscopic feedback, leads to the following equation for the dynamics of the DM  particles
\be
\frac{dp^{\alpha}}{d\tau}=\left(\nabla_\beta p^\alpha\right) u^\beta = m_\phi \beta \partial^\alpha \varphi\label{mconserv}\,.
\ee
where $p^{\alpha}=m_\phi u^\alpha$, and $m_\phi$, $\tau$ and $u^\alpha$ are, respectively,  the mass, the proper time and the coordinates of the $N+1$-velocity of the DM particle (here $N$ represents the number of spatial dimensions).

\subsection{Dynamics in $1+1$ dimensions}

In this paper, we investigate the dynamics of DM particles non-minimally coupled to DE, focusing on microscopic feedback with a  characteristic timescale much smaller than the cosmological timescale. Therefore, in the following, we shall neglect the expansion of the Universe.  In a $1+1$ dimensional Minkowski space-time the line element can be written as $ds^2=-dt^2+dz^2$ so that the equations of motion for the scalar fields $\phi$ and $\varphi$ are given respectively by
\bq
{\ddot \phi} - \phi''&=& -\frac{\partial U}{\partial \phi}\,, \label{phieqmM}\\
{\ddot \varphi} - \varphi''&=& -\frac{d W}{d\varphi}+2\beta U \label{varphieqmE}\,,
\eq
where a dot denotes a derivative with respect to the physical time $t$ and a prime represents a derivative with respect to the space coordinate $z$.

The components of the energy-momentum tensor of the DM and DE components can now be written as
\bq
T^{00}_{DM}&=&\frac{{\dot \phi}^2}{2}+\frac{\phi'^2}{2}+U(\phi,\varphi)\,,\\
T^{0z}_{DM}&=&-{\dot \phi}\phi'\,,\\
T^{zz}_{DM}&=&\frac{{\dot \phi}^2}{2}+\frac{\phi'^2}{2}-U(\phi,\varphi)\,,\\
T^{00}_{DE}&=&\frac{{\dot \varphi}^2}{2}+\frac{\varphi'^2}{2}+W(\varphi)\,,\\
T^{0z}_{DE}&=&-{\dot \varphi}\varphi'\,,\\
T^{zz}_{DE}&=&\frac{{\dot \varphi}^2}{2}+\frac{\varphi'^2}{2}-W(\varphi)\,,
\eq

Let us start by considering a static DM particle, and assume that $\phi=\phi(z)$ and $\lambda={\rm const}$. In this case Eq. (\ref{phieqmM}) becomes
\be
 \phi''= \frac{dU}{d\phi} \label{phieqmM1}\,,
\ee
and it can be integrated to give
\be
\frac{\phi'^2}{2} = U\,,\label{KeqU}
\ee
assuming that $|\phi| \to \eta$ for $z \to \pm \infty$. Eq. (\ref{KeqU}) implies that in the static case the total energy density is equal to $2U$. If the particle is located at $z=0$,  Eq. (\ref{phieqmM1}) has the following solution
\be
\phi = \pm \eta \tanh\left(\frac{z}{{\sqrt 2}R}\right)\,,
\ee
with
\be
R=\lambda^{-1/2} \eta^{-1}\,.
\ee
The rest mass of the particle is given by
\bq
m_\phi&=&\int_{- \infty}^{\infty} T^{00} dz = 2 \int_{- \infty}^{\infty} U dz = \frac{8 {\sqrt 2}}{3} U_{max} R  = \nonumber \\
&=& \frac{2{\sqrt 2}}{3}\lambda^{1/2}  \eta^3\,,
\eq
which implies that
\be
\beta\equiv-\frac12\frac{d \ln \lambda}{d\varphi}=-\frac{d \ln m_\phi}{d\varphi}\label{beta}\,.
\ee
Here $U_{max} \equiv U(\phi=0) = \lambda \eta^4/4$.

The evolution of the total energy $E=m_\phi \gamma$ and linear momentum $p=m_\phi \gamma v$ associated with the DM particles may be calculated using Eqs. (\ref{TDM}) and (\ref{DMQ})
\bq
\frac{dE}{dt}&=& \int \partial_0 T^{00}_{DM} dz = \nonumber\\
&=& 2\int\beta U(\phi,\varphi) \partial^0 \varphi dz \sim - \frac{\beta m_\phi  {\dot \varphi}}{\gamma}\,,\label{dEdt}\\
F=\frac{dp}{dt} &=& -\int \partial_0 T^{0z}_{DM} dz = \nonumber\\
&=&2\int\beta U(\phi,\varphi) \partial^z \varphi dz \sim \frac{\beta m_\phi \varphi'}{\gamma}\label{dpdt}\,,
\eq
where $v$ is the DM particle velocity and $\gamma \equiv (1-v^2)^{-1/2}$. Here, we assume that ${\dot \varphi}$ and $\varphi'$ are smooth on the characteristic length scale of the particle, which is not the case in general. Eq. (\ref{dpdt}) implies that, in the absence of gradients of the DE field $\varphi$, the DM particles do conserve their linear momentum. The gradients of the field are responsible for a fifth force mediated by the DE field $\varphi$ which, in the Newtonian limit, is $\beta^2/(4\pi G)$ times stronger than gravity. Taking into account that $p^0=E$, $p^z=p$, and $dt/d\tau = \gamma$, it is simple to show that the covariant form of Eqs. (\ref{dEdt}) and (\ref{dpdt}) is given by Eq. (\ref{mconserv}).

\section{Numerical simulations\label{sec3}}

In this section we describe the results of field theory numerical simulations in $1+1$ dimensions of the dynamics of DM particles interacting non-minimally with DE. The equations of motion (\ref{phieqmM}) and (\ref{varphieqmE}) are solved using a 2nd order Runge-Kutta time-integration method combined with a 2nd order finite difference scheme for the spatial derivatives.
In all the simulations we assume that 
\be
\beta=\beta_{on} \Theta(\varphi-\varphi_{on})-\beta_{on} \Theta(\varphi-\varphi_{off})\label{beta1}\,,
\ee
where $\Theta$ is the Heaviside function. This implies that $\beta$ is switched on to $\beta=\beta_{on}$ for $\varphi=\varphi_{on}$ and switched off for $\varphi=\varphi_{off}$. Using Eqs. (\ref{beta}) and (\ref{beta1}) one finds that
\be
\beta_{on}=-\frac{1}{\varphi_{off}-\varphi_{on}}\ln\left(\frac{m_{\phi} (\varphi_{off})}{m_{\phi} (\varphi_{on})}\right)\,.
\ee

Our simulations run in a one-dimensional grid of size $L=1 \times 10^4$ with periodic boundary conditions. The initial conditions for the scalar fields ($\phi$ and $\varphi$) and their time derivatives are given by
\bq
\phi_i &=& \eta \left(\tanh\left(\frac{\gamma_i(z-L/4)}{{\sqrt 2} R_i}\right)\right.- \nonumber\\
&-&\left.\tanh\left(\frac{ \gamma_i(z-3L/4)}{{\sqrt 2} R_i}\right)-1\right)\,,\\
{\dot \phi}_i&=&\pm v_i \gamma_i {\sqrt {2 U (\phi_i,\varphi_i)}}\label{dotphii}\,,\\
\varphi_i  &=& 0\,,\\
{\dot \varphi_i} &=& 4 \times 10^{-3}\,,
\eq
where $v_i=1 \times 10^{-3}$, $R_i=20$, the subscript `i' refers to the initial time $t_i=0$, and the $+$ sign in Eq. (\ref{dotphii}) applies for $z>L/2$, while the $-$ sign applies for $z<L/2$. The initial conditions for $\phi$ and ${\dot \phi}$ describe a DM particle and an antiparticle at a distance $L/2 \gg R$ from one another, moving both with velocity $v=v_i$. The final time $t=t_f=5 \times 10^3$ is such that the particle and the anti-particle are effectively isolated from each other during the course of the simulation (the fifth force between them, mediated by the DE scalar field, would only be felt for $t>t_f$).

In our simulations we take
\be
\xi \equiv m_{\phi} (\varphi_{off})/m_{\phi} (\varphi_{on})\,,
\ee
to be either $\xi=1.2$ or $0.8$, corresponding to a model where the mass of the DM particles increases or decreases by $20 \%$, respectively. Our units of velocity and energy are such that $c=1$ and $\eta=1$ (except for these conditions the choice of units is arbitrary). For simplicity we shall also assume that $W={\rm const}$.

\begin{figure}
\includegraphics[width=3.4in]{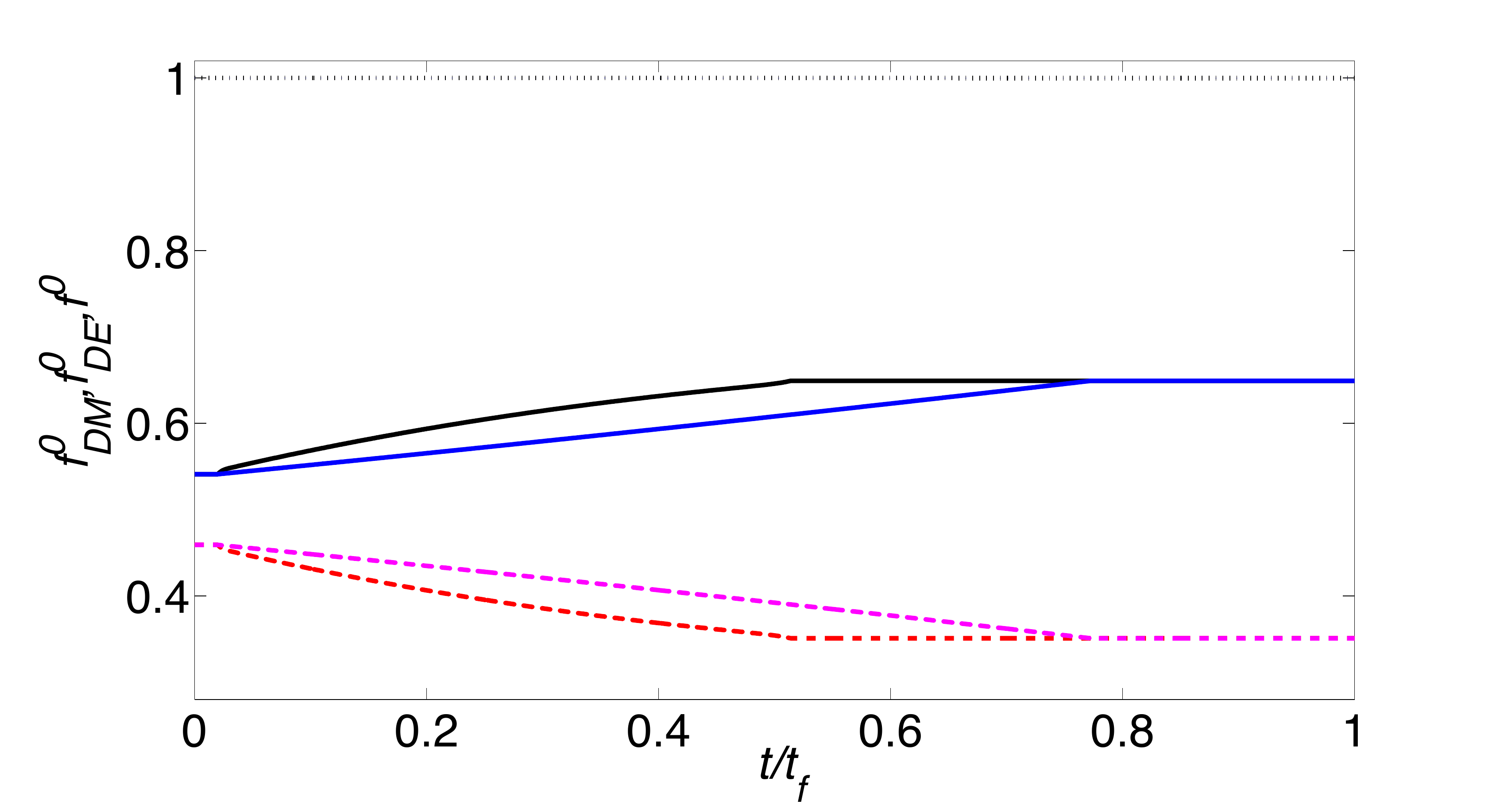}
\includegraphics[width=3.4in]{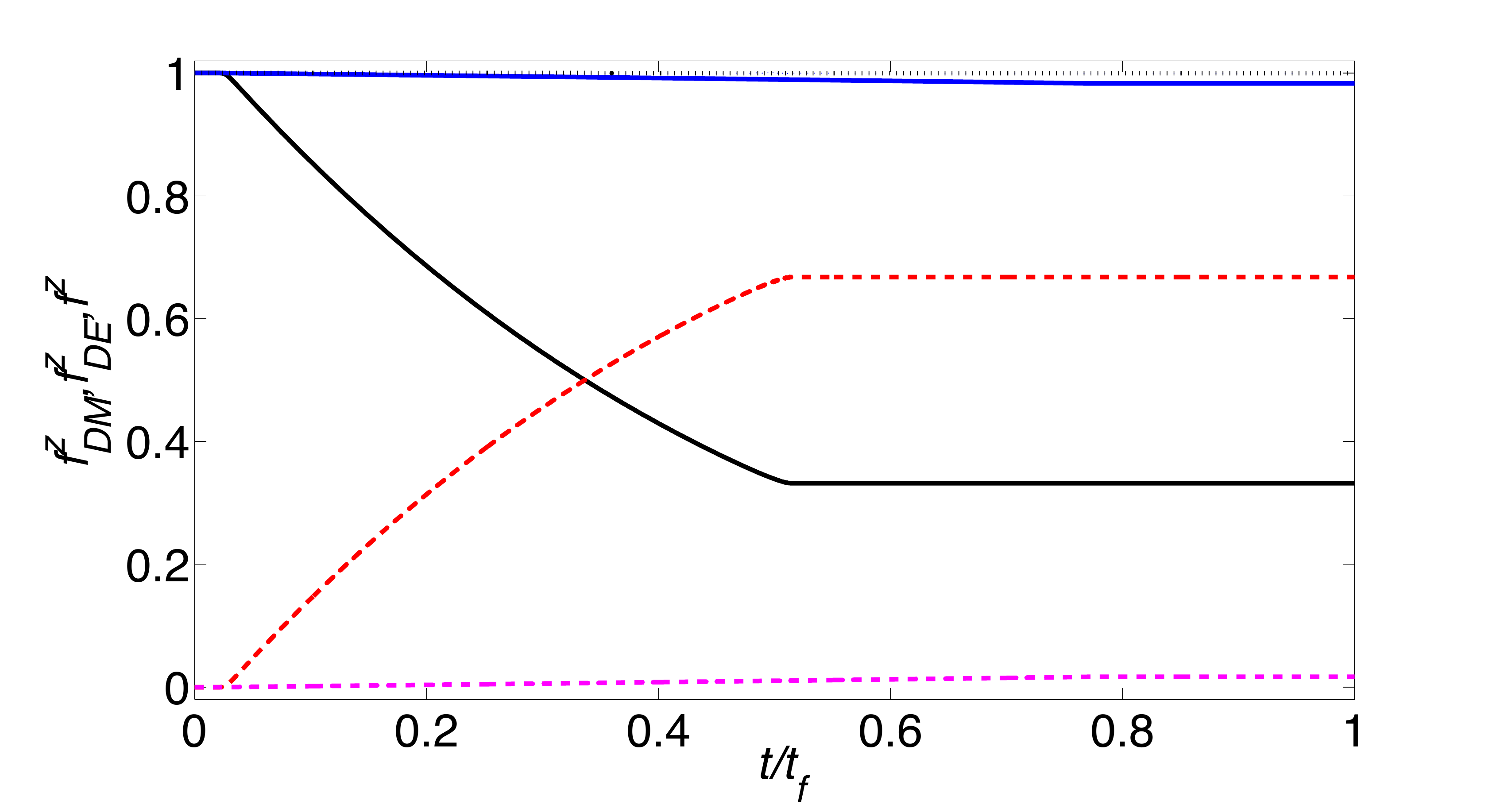}
\includegraphics[width=3.4in]{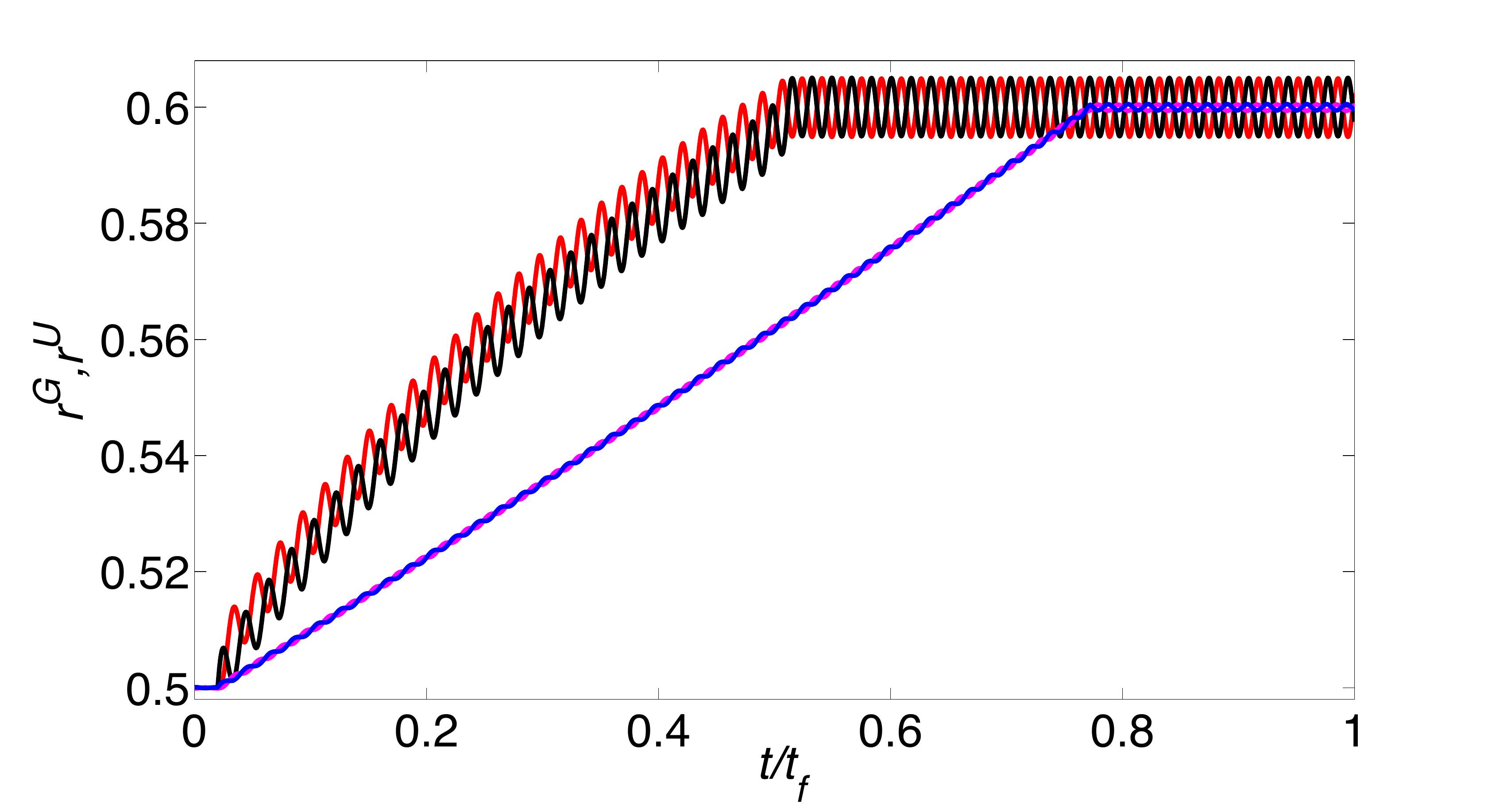}
\caption{The top and middle panels show, respectively,  the evolution of $f^0_{DM}$ and  $f^z_{DM}$ (solid lines) and of $f^0_{DE}$ and $f^z_{DE}$ (dashed lines) considering $\xi=1.2$ for $\alpha=2$ (black and red lines) and $\alpha=0.2$ (blue and magenta lines). Note that the values of $f^0$ and $f^z$, represented by dotted lines in the top and middle panels, respectively, are always very close to unity as required by energy conservation. The lower panel shows the evolution of $r^U$ and  $r^G$ for $\alpha=2$ (black and red lines, respectively) and $\alpha=0.2$ (blue and magenta lines, respectively).}
\label{Fig1}
\end{figure}

\begin{figure}
\includegraphics[width=3.4in]{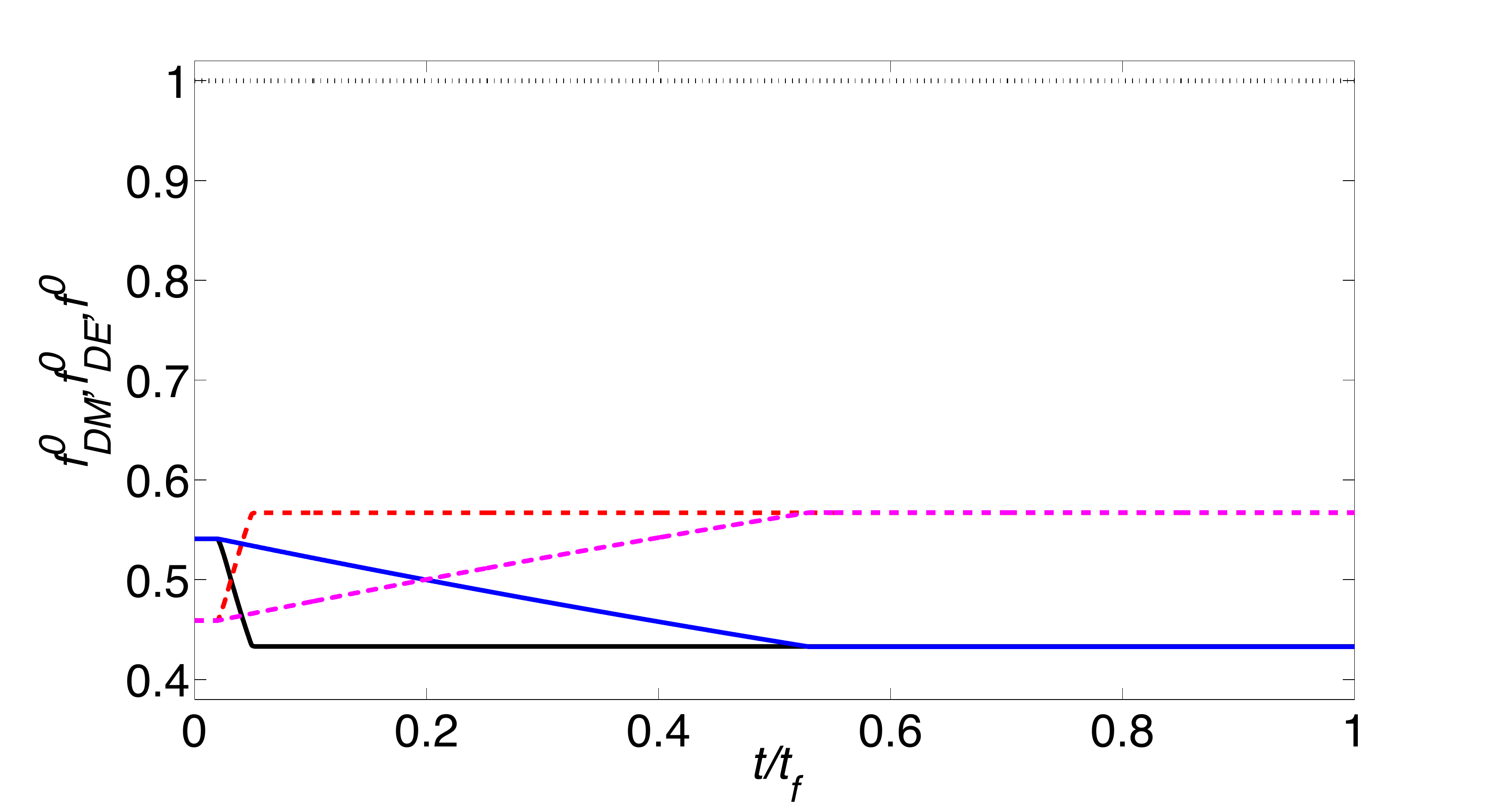}
\includegraphics[width=3.4in]{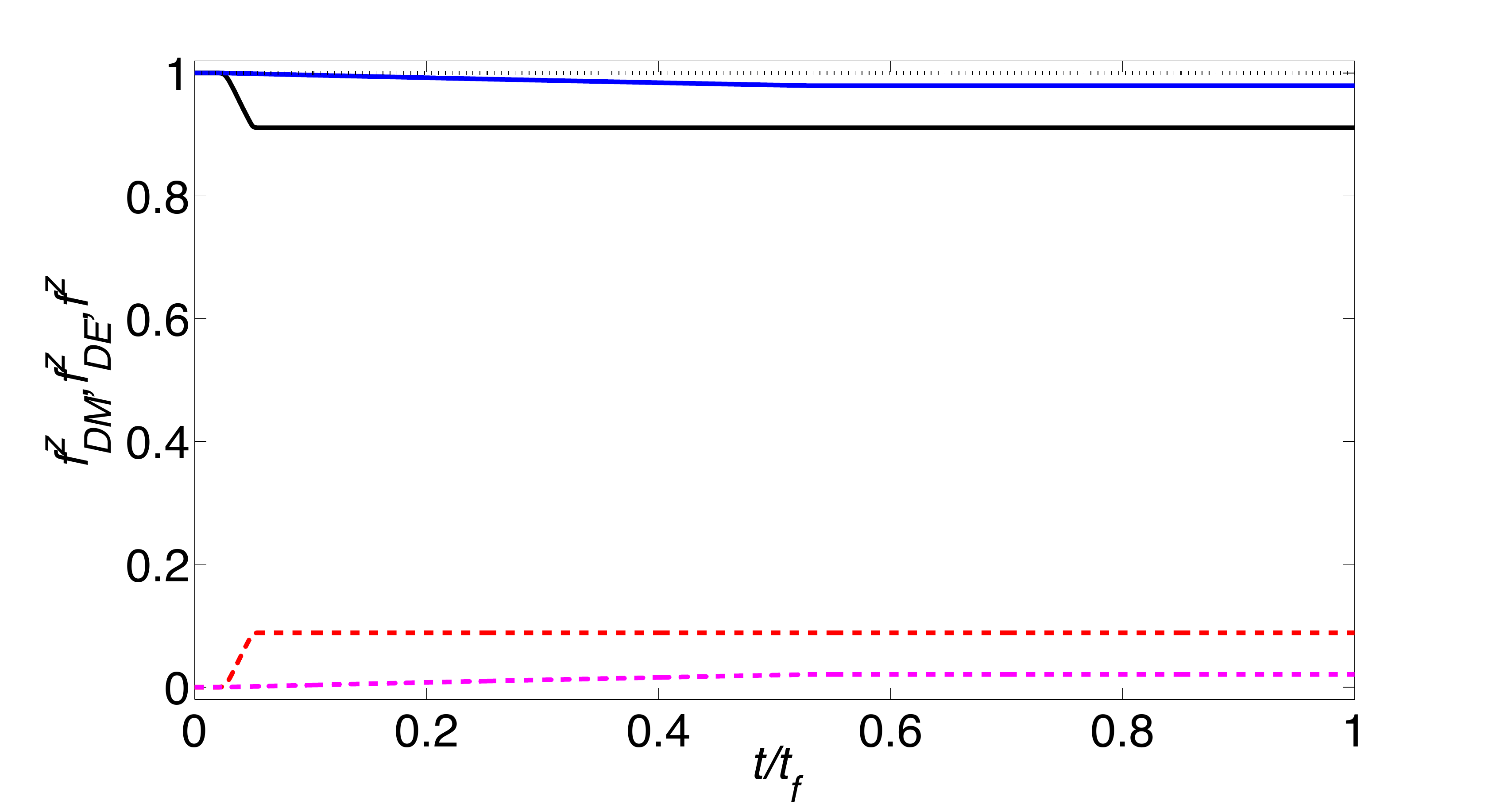}
\includegraphics[width=3.4in]{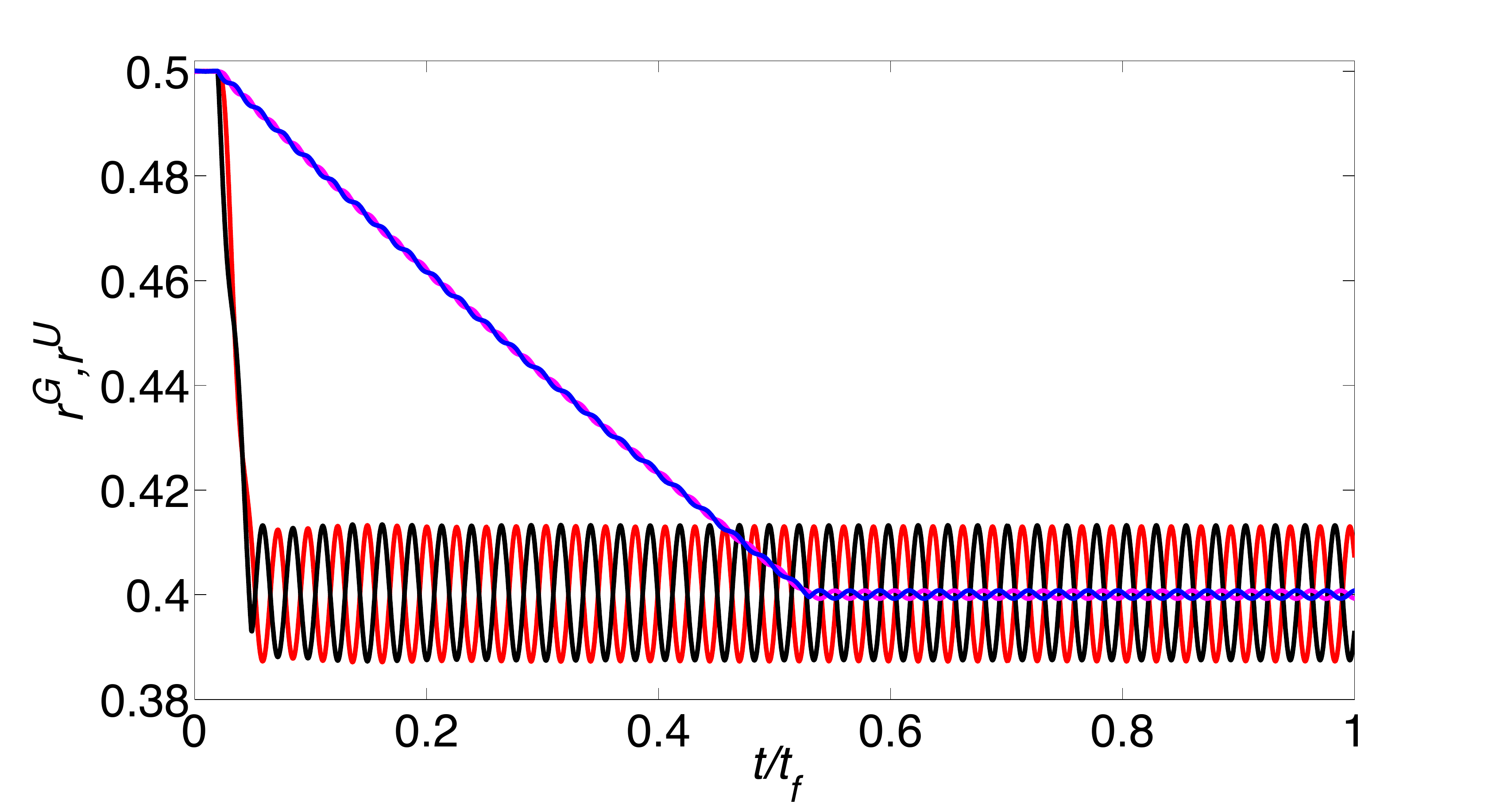}
\caption{Analogous to Fig. \ref{Fig1} for $\xi=0.8$.}
\label{Fig2}
\end{figure}

Let us define the parameter $\alpha$ by
\be
\alpha \equiv \frac{(\lambda_{off}-\lambda_{on}) \eta^4}{2{\dot \varphi_i}^2} \times \frac{R_i}{\Delta t_\varphi} \,,
\ee
where $\Delta t_\varphi$ is a characteristic time associated to the variation of $\varphi$ defined as
\be
\Delta t_\varphi \equiv \frac{\varphi_{off}-\varphi_{on}}{\dot \varphi_i} \,.
\ee
In our simulations we set $\varphi_{on}=1 \times 10^2 {\dot \varphi_i}$ and $\varphi_{off}$ is determined by the value of $\alpha$, which we choose to be either $0.2$ or $2$. The parameter $\alpha$ represents the ratio between the average variation of the characteristic microscopic energy density of the particle on a timescale equal to $R_i$, 
\be 
\Delta U_{max} R_i/(\Delta t_\varphi)\,, \nonumber
\ee 
and the initial kinetic energy density associated with the DE, ${\dot \varphi_i}^2/2$. For $\alpha \gsim 1$ microscopic feedback effects associated with the evolution of the mass of the DM particles are expected to be large. On the other hand, for $\alpha \ll 1$ these effects are expected to be small and the standard results for the dynamics of DM particles non-minimally coupled to the DE field should hold. 

The following functions
\bq
f^0_{DM} &\equiv& \frac{\int T^{00}_{DM} \, dz}{\int T^{00}_{t=0} \, dz}\,, \quad f^0_{DE} \equiv \frac{\int T^{00}_{DE} \, dz}{\int T^{00}_{t=0} \, dz}\,,\\ 
f^z_{DM} &\equiv& \frac{\int T^{0z}_{DM} \, dz}{\int T^{0z}_{t=0} \, dz}\,, \quad f^z_{DE} \equiv \frac{\int T^{0z}_{DE} \, dz}{\int T^{0z}_{t=0} \, dz}\,,\\
f^0&=&f^0_{DM}+f^0_{DE}\,, \quad f^z=f^z_{DM}+f^z_{DE}\,,\\
\,r^G &\equiv&  \frac{\int  \phi'^2 \, dz}{2\int T^{00}_{DM|t=0} \, dz}\,, \quad \,r^U \equiv  \frac{\int U \, dz}{\int T^{00}_{DM|t=0} \, dz}\,,
\eq
were computed with our simulations for $\xi=1.2$ (Fig. \ref{Fig1}) and $\xi=0.8$ (Fig. \ref{Fig2}). Here we have disregarded the contribution of the DE potential $W(\varphi)$ in the calculation of $T^{00}_{DE}$ and $T^{00}$.

Fig. \ref{Fig1} (top panel) shows the evolution of $f^0_{DM}$ (solid lines) and  $f^0_{DE}$ (dashed lines) for $\alpha=2$ (black and red lines) and $\alpha=0.2$ (blue and magenta lines). It shows the increase of $f^0_{DM}$ by $20 \%$, associated to the increase of the mass of the DM particles, as well as the corresponding decrease of $f^0_{DE}$ by the same amount. The evolution is faster in the  $\alpha=2$ case than for $\alpha=0.2$. The value of $f^0$ (dotted line) is always very close to unity, both in the $\alpha=2$ and in the $\alpha=0.2$ cases, as required by energy conservation.

Fig. \ref{Fig1} (middle panel) shows the evolution of $f^z_{DM}$ (solid lines) and  $f^z_{DE}$ (dashed lines) for $\alpha=2$ (black and red lines) and $\alpha=0.2$ (blue and magenta lines). It shows that the decrease of $f^z_{DM}$, associated to the decrease of the linear momentum of the DM particles, is exactly compensated by the increase of $f^z_{DE}$, so that $f^z$ (dotted line) is always very close to unity, as required by linear momentum conservation. For $\alpha=2$ the linear momentum of the DM particles is changed dramatically while for $\alpha=0.2$ it is nearly conserved. This is not surprising, since microscopic feedback should have a negligible impact on the dynamics of the DM particles for $\alpha \ll 1$, only becoming significant for  $\alpha \gsim 1$.

Fig. \ref{Fig1} (lower panel) shows the evolution of $r^U$ and  $r^G$ for $\alpha=2$ (black and red lines, respectively) and $\alpha=0.2$ (blue and magenta lines, respectively). It shows that $r^U$ and $r^G$ have always a similar magnitude (note that $r^U = r^G$ in the static case) and present long lived oscillations (more pronounced for $\alpha=2$ than for $\alpha=0.2$). Notice that $r^G$ and $r^U$ have a phase lag of $\pi$ and that $r^U+r^G$ evolves from approximately $1$ to $1.2$ as a direct consequence of the $20\%$ decrease in the mass of the DM particles. The DM energy associated with the time derivative of $\phi$ increases with $\alpha$ but it is very small both for $\alpha=0.2$ and $\alpha=2$.

Fig. \ref{Fig2} is analogous to Fig. \ref{Fig1} but now for $\xi=0.8$, meaning that the mass of the DM particles decreases by $20\%$. One of the most significant differences between Figs. \ref{Fig1} and \ref{Fig2} is related to the very different timescales for the energy-momentum transfer between the DM particles and the DE field for $\alpha=2$, which happens even though the value of $\Delta t_\varphi$ is the same for $\xi=1.2$ and $\xi=0.8$. This has to do with the fact that a rapid  increase (decrease) of the mass of the DM particles is also associated to a rapid decrease (increase) of the kinetic energy of the DE scalar field, thus leading to a switch off time significantly larger (smaller) than $\Delta t_\varphi$ for $\xi=1.2$ ($\xi=0.8$) in the $\alpha=2$ case. Hence, for $\alpha \gsim 1$ the microscopic and macroscopic values of $\beta$ can be in general be very different in models where $\beta = \beta(\varphi)$.

\section{\label{conc} Discussion and conclusions}

In this paper we have investigated the classical microscopic dynamics of DM particles non-minimally coupled to a DE field, using a simple model which is expected to capture the essential features of generic scenarios where the mass of the DM particles is coupled to the DE scalar field. This analysis complements previous studies where the coupling between DM and DE energy has been considered only at a macroscopic level, and it should eventually be validated at a quantum level. We have shown using analytical arguments and numerical simulations that although the standard results for the dynamics of the DM particles can be recovered if the space-time variation of the DE scalar field is sufficiently smooth on the characteristic length scale of the DM particles, in general the DE coupling introduces non-trivial backreaction effects. 

We have found that microscopic feedback becomes important when the variation of the characteristic microscopic energy density of the particle on a timescale equal to $R$ is of the order of kinetic energy density associated with the DE. In three spatial dimensions this is equivalent to the condition
\be
 \frac{\beta {\dot \varphi} m_\phi}{R^2}  \sim \rho_\varphi (w_\varphi+1) \,,
\ee
where $\rho_\varphi$ and $p_\varphi = w_\varphi \rho_\varphi$ are the DE energy density and pressure. Hence, microscopic feedback should be taken into account for DM masses
\be
m_\phi  \gsim m_* = \frac{\rho_\varphi (w_\varphi+1) R^2}{\beta {\dot \varphi}}\,.
\ee
As a reference, a DM particle with a radius and mass similar to that of the proton ($m_p \sim 10^{-27} \, {\rm kg}$ and $r_p \sim 10^{-15} \, {\rm m}$) microscopic feedback would be important today even for a value of $\beta_0$ as small as $\beta_0 {\dot \varphi}_0 \sim 10^{-4} H_0$ (here a `0' refers to the present time and we have assumed that $|w_{\varphi 0}-1| \lsim 0.1$). Note also that, $m_* \to 0$ for $R \to 0$, irrespectively of the value of $\beta$. 

We have demonstrated that for $m_\phi \gsim m_*$ a significant transfer of linear momentum between the DM particles and the DE scalar field is associated to microscopic feedback, which could have an important impact on the growth of cosmic structures. On the other hand, for $m_\phi \ll m_*$ the linear momentum transfer due to microscopic feedback is negligible. We have also found that for $m_\phi \gsim m_*$ the microscopic energy and linear momentum DE density around the DM particles can in general be very different from the average  ones. This implies that microscopic feedback may play a relevant role on the space-time evolution of the DE equation of state, and can be associated with very different microscopic and macroscopic values of $\beta$ in models where $\beta=\beta(\varphi)$. The requirement that microscopic feedback effects do not have a dramatic impact on the dynamics of both DM and DE may turn out to be one of the strongest constraints on the nature of DM particles in coupled dark energy scenarios.

\begin{acknowledgments}

This work was supported by Funda{\c c}\~ao para a Ci\^encia e a Tecnologia (FCT) through the Investigador FCT contract of reference IF/00863/2012 and POPH/FSE (EC) by FEDER funding through the program "Programa Operacional de Factores de Competitividade - COMPETE.
\end{acknowledgments}


\bibliography{IDE}

\begin{thebibliography}{30}
\expandafter\ifx\csname natexlab\endcsname\relax\def\natexlab#1{#1}\fi
\expandafter\ifx\csname bibnamefont\endcsname\relax
  \def\bibnamefont#1{#1}\fi
\expandafter\ifx\csname bibfnamefont\endcsname\relax
  \def\bibfnamefont#1{#1}\fi
\expandafter\ifx\csname citenamefont\endcsname\relax
  \def\citenamefont#1{#1}\fi
\expandafter\ifx\csname url\endcsname\relax
  \def\url#1{\texttt{#1}}\fi
\expandafter\ifx\csname urlprefix\endcsname\relax\def\urlprefix{URL }\fi
\providecommand{\bibinfo}[2]{#2}
\providecommand{\eprint}[2][]{\url{#2}}

\bibitem[{\citenamefont{Suzuki et~al.}(2012)\citenamefont{Suzuki, Rubin,
  Lidman, Aldering, Amanullah et~al.}}]{Suzuki:2011hu}
\bibinfo{author}{\bibfnamefont{N.}~\bibnamefont{Suzuki}},
  \bibinfo{author}{\bibfnamefont{D.}~\bibnamefont{Rubin}},
  \bibinfo{author}{\bibfnamefont{C.}~\bibnamefont{Lidman}},
  \bibinfo{author}{\bibfnamefont{G.}~\bibnamefont{Aldering}},
  \bibinfo{author}{\bibfnamefont{R.}~\bibnamefont{Amanullah}},
  \bibnamefont{et~al.}, \bibinfo{journal}{Astrophys.J.}
  \textbf{\bibinfo{volume}{746}}, \bibinfo{pages}{85} (\bibinfo{year}{2012}).

\bibitem[{\citenamefont{Anderson et~al.}(2013)\citenamefont{Anderson, Aubourg,
  Bailey, Bizyaev, Blanton et~al.}}]{Anderson:2012sa}
\bibinfo{author}{\bibfnamefont{L.}~\bibnamefont{Anderson}},
  \bibinfo{author}{\bibfnamefont{E.}~\bibnamefont{Aubourg}},
  \bibinfo{author}{\bibfnamefont{S.}~\bibnamefont{Bailey}},
  \bibinfo{author}{\bibfnamefont{D.}~\bibnamefont{Bizyaev}},
  \bibinfo{author}{\bibfnamefont{M.}~\bibnamefont{Blanton}},
  \bibnamefont{et~al.}, \bibinfo{journal}{Mon.Not.Roy.Astron.Soc.}
  \textbf{\bibinfo{volume}{427}}, \bibinfo{pages}{3435} (\bibinfo{year}{2013}).

\bibitem[{\citenamefont{Parkinson et~al.}(2012)\citenamefont{Parkinson,
  Riemer-Sorensen, Blake, Poole, Davis et~al.}}]{Parkinson:2012vd}
\bibinfo{author}{\bibfnamefont{D.}~\bibnamefont{Parkinson}},
  \bibinfo{author}{\bibfnamefont{S.}~\bibnamefont{Riemer-Sorensen}},
  \bibinfo{author}{\bibfnamefont{C.}~\bibnamefont{Blake}},
  \bibinfo{author}{\bibfnamefont{G.~B.} \bibnamefont{Poole}},
  \bibinfo{author}{\bibfnamefont{T.~M.} \bibnamefont{Davis}},
  \bibnamefont{et~al.}, \bibinfo{journal}{Phys.Rev.}
  \textbf{\bibinfo{volume}{D86}}, \bibinfo{pages}{103518}
  (\bibinfo{year}{2012}).

\bibitem[{\citenamefont{Hinshaw et~al.}(2013)}]{Hinshaw:2012aka}
\bibinfo{author}{\bibfnamefont{G.}~\bibnamefont{Hinshaw}} \bibnamefont{et~al.}
  (\bibinfo{collaboration}{WMAP}), \bibinfo{journal}{Astrophys.J.Suppl.}
  \textbf{\bibinfo{volume}{208}}, \bibinfo{pages}{19} (\bibinfo{year}{2013}).

\bibitem[{\citenamefont{Ade et~al.}(2014)}]{Ade:2013zuv}
\bibinfo{author}{\bibfnamefont{P.~A.~R.} \bibnamefont{Ade}}
  \bibnamefont{et~al.} (\bibinfo{collaboration}{Planck Collaboration}),
  \bibinfo{journal}{Astron.Astrophys.} \textbf{\bibinfo{volume}{571}},
  \bibinfo{pages}{A16} (\bibinfo{year}{2014}).

\bibitem[{\citenamefont{Ade et~al.}(2015{\natexlab{a}})}]{Planck:2015xua}
\bibinfo{author}{\bibfnamefont{P.~A.~R.} \bibnamefont{Ade}}
  \bibnamefont{et~al.} (\bibinfo{collaboration}{Planck Collaboration})
  (\bibinfo{year}{2015}{\natexlab{a}}), \eprint{1502.01589}.

\bibitem[{\citenamefont{Copeland et~al.}(2006)\citenamefont{Copeland, Sami, and
  Tsujikawa}}]{Copeland:2006wr}
\bibinfo{author}{\bibfnamefont{E.~J.} \bibnamefont{Copeland}},
  \bibinfo{author}{\bibfnamefont{M.}~\bibnamefont{Sami}}, \bibnamefont{and}
  \bibinfo{author}{\bibfnamefont{S.}~\bibnamefont{Tsujikawa}},
  \bibinfo{journal}{Int. J. Mod. Phys.} \textbf{\bibinfo{volume}{D15}},
  \bibinfo{pages}{1753} (\bibinfo{year}{2006}).

\bibitem[{\citenamefont{Frieman et~al.}(2008)\citenamefont{Frieman, Turner, and
  Huterer}}]{Frieman:2008sn}
\bibinfo{author}{\bibfnamefont{J.}~\bibnamefont{Frieman}},
  \bibinfo{author}{\bibfnamefont{M.}~\bibnamefont{Turner}}, \bibnamefont{and}
  \bibinfo{author}{\bibfnamefont{D.}~\bibnamefont{Huterer}},
  \bibinfo{journal}{Ann. Rev. Astron. Astrophys.}
  \textbf{\bibinfo{volume}{46}}, \bibinfo{pages}{385} (\bibinfo{year}{2008}).

\bibitem[{\citenamefont{Caldwell and Kamionkowski}(2009)}]{Caldwell:2009ix}
\bibinfo{author}{\bibfnamefont{R.~R.} \bibnamefont{Caldwell}} \bibnamefont{and}
  \bibinfo{author}{\bibfnamefont{M.}~\bibnamefont{Kamionkowski}},
  \bibinfo{journal}{Ann. Rev. Nucl. Part. Sci.} \textbf{\bibinfo{volume}{59}},
  \bibinfo{pages}{397} (\bibinfo{year}{2009}).

\bibitem[{\citenamefont{Li et~al.}(2011)\citenamefont{Li, Li, Wang, and
  Wang}}]{Li:2011sd}
\bibinfo{author}{\bibfnamefont{M.}~\bibnamefont{Li}},
  \bibinfo{author}{\bibfnamefont{X.-D.} \bibnamefont{Li}},
  \bibinfo{author}{\bibfnamefont{S.}~\bibnamefont{Wang}}, \bibnamefont{and}
  \bibinfo{author}{\bibfnamefont{Y.}~\bibnamefont{Wang}},
  \bibinfo{journal}{Commun.Theor.Phys.} \textbf{\bibinfo{volume}{56}},
  \bibinfo{pages}{525} (\bibinfo{year}{2011}).

\bibitem[{\citenamefont{Bamba et~al.}(2012)\citenamefont{Bamba, Capozziello,
  Nojiri, and Odintsov}}]{Bamba:2012cp}
\bibinfo{author}{\bibfnamefont{K.}~\bibnamefont{Bamba}},
  \bibinfo{author}{\bibfnamefont{S.}~\bibnamefont{Capozziello}},
  \bibinfo{author}{\bibfnamefont{S.}~\bibnamefont{Nojiri}}, \bibnamefont{and}
  \bibinfo{author}{\bibfnamefont{S.~D.} \bibnamefont{Odintsov}},
  \bibinfo{journal}{Astrophys.Space Sci.} \textbf{\bibinfo{volume}{342}},
  \bibinfo{pages}{155} (\bibinfo{year}{2012}).

\bibitem[{\citenamefont{Wetterich}(1995)}]{Wetterich:1994bg}
\bibinfo{author}{\bibfnamefont{C.}~\bibnamefont{Wetterich}},
  \bibinfo{journal}{Astron. Astrophys.} \textbf{\bibinfo{volume}{301}},
  \bibinfo{pages}{321} (\bibinfo{year}{1995}).

\bibitem[{\citenamefont{Amendola}(2000)}]{Amendola:1999er}
\bibinfo{author}{\bibfnamefont{L.}~\bibnamefont{Amendola}},
  \bibinfo{journal}{Phys. Rev.} \textbf{\bibinfo{volume}{D62}},
  \bibinfo{pages}{043511} (\bibinfo{year}{2000}).

\bibitem[{\citenamefont{Zimdahl and Pavon}(2001)}]{Zimdahl:2001ar}
\bibinfo{author}{\bibfnamefont{W.}~\bibnamefont{Zimdahl}} \bibnamefont{and}
  \bibinfo{author}{\bibfnamefont{D.}~\bibnamefont{Pavon}},
  \bibinfo{journal}{Phys.Lett.} \textbf{\bibinfo{volume}{B521}},
  \bibinfo{pages}{133} (\bibinfo{year}{2001}).

\bibitem[{\citenamefont{Farrar and Peebles}(2004)}]{Farrar:2003uw}
\bibinfo{author}{\bibfnamefont{G.~R.} \bibnamefont{Farrar}} \bibnamefont{and}
  \bibinfo{author}{\bibfnamefont{P.~E.} \bibnamefont{Peebles}},
  \bibinfo{journal}{Astrophys.J.} \textbf{\bibinfo{volume}{604}},
  \bibinfo{pages}{1} (\bibinfo{year}{2004}).

\bibitem[{\citenamefont{Gumjudpai et~al.}(2005)\citenamefont{Gumjudpai, Naskar,
  Sami, and Tsujikawa}}]{Gumjudpai:2005ry}
\bibinfo{author}{\bibfnamefont{B.}~\bibnamefont{Gumjudpai}},
  \bibinfo{author}{\bibfnamefont{T.}~\bibnamefont{Naskar}},
  \bibinfo{author}{\bibfnamefont{M.}~\bibnamefont{Sami}}, \bibnamefont{and}
  \bibinfo{author}{\bibfnamefont{S.}~\bibnamefont{Tsujikawa}},
  \bibinfo{journal}{JCAP} \textbf{\bibinfo{volume}{0506}}, \bibinfo{pages}{007}
  (\bibinfo{year}{2005}).

\bibitem[{\citenamefont{Pettorino and Baccigalupi}(2008)}]{Pettorino:2008ez}
\bibinfo{author}{\bibfnamefont{V.}~\bibnamefont{Pettorino}} \bibnamefont{and}
  \bibinfo{author}{\bibfnamefont{C.}~\bibnamefont{Baccigalupi}},
  \bibinfo{journal}{Phys. Rev.} \textbf{\bibinfo{volume}{D77}},
  \bibinfo{pages}{103003} (\bibinfo{year}{2008}).

\bibitem[{\citenamefont{Caldera-Cabral
  et~al.}(2009{\natexlab{a}})\citenamefont{Caldera-Cabral, Maartens, and
  Urena-Lopez}}]{CalderaCabral:2008bx}
\bibinfo{author}{\bibfnamefont{G.}~\bibnamefont{Caldera-Cabral}},
  \bibinfo{author}{\bibfnamefont{R.}~\bibnamefont{Maartens}}, \bibnamefont{and}
  \bibinfo{author}{\bibfnamefont{L.~A.} \bibnamefont{Urena-Lopez}},
  \bibinfo{journal}{Phys. Rev.} \textbf{\bibinfo{volume}{D79}},
  \bibinfo{pages}{063518} (\bibinfo{year}{2009}{\natexlab{a}}).

\bibitem[{\citenamefont{Caldera-Cabral
  et~al.}(2009{\natexlab{b}})\citenamefont{Caldera-Cabral, Maartens, and
  Schaefer}}]{CalderaCabral:2009ja}
\bibinfo{author}{\bibfnamefont{G.}~\bibnamefont{Caldera-Cabral}},
  \bibinfo{author}{\bibfnamefont{R.}~\bibnamefont{Maartens}}, \bibnamefont{and}
  \bibinfo{author}{\bibfnamefont{B.~M.} \bibnamefont{Schaefer}},
  \bibinfo{journal}{JCAP} \textbf{\bibinfo{volume}{0907}}, \bibinfo{pages}{027}
  (\bibinfo{year}{2009}{\natexlab{b}}).

\bibitem[{\citenamefont{Baldi}(2011)}]{Baldi:2010pq}
\bibinfo{author}{\bibfnamefont{M.}~\bibnamefont{Baldi}}, \bibinfo{journal}{Mon.
  Not. Roy. Astron. Soc.} \textbf{\bibinfo{volume}{414}}, \bibinfo{pages}{116}
  (\bibinfo{year}{2011}).

\bibitem[{\citenamefont{Ade et~al.}(2015{\natexlab{b}})}]{Ade:2015yua}
\bibinfo{author}{\bibfnamefont{P.~A.~R.} \bibnamefont{Ade}}
  \bibnamefont{et~al.} (\bibinfo{collaboration}{Planck Collaboration})
  (\bibinfo{year}{2015}{\natexlab{b}}), \eprint{1502.01590}.

\bibitem[{\citenamefont{Avelino and da~Silva}(2012)}]{Avelino:2012tc}
\bibinfo{author}{\bibfnamefont{P.}~\bibnamefont{Avelino}} \bibnamefont{and}
  \bibinfo{author}{\bibfnamefont{H.}~\bibnamefont{da~Silva}},
  \bibinfo{journal}{Phys.Lett.} \textbf{\bibinfo{volume}{B714}},
  \bibinfo{pages}{6} (\bibinfo{year}{2012}), \eprint{1201.0550}.

\bibitem[{\citenamefont{Salvatelli et~al.}(2013)\citenamefont{Salvatelli,
  Marchini, Lopez-Honorez, and Mena}}]{Salvatelli:2013wra}
\bibinfo{author}{\bibfnamefont{V.}~\bibnamefont{Salvatelli}},
  \bibinfo{author}{\bibfnamefont{A.}~\bibnamefont{Marchini}},
  \bibinfo{author}{\bibfnamefont{L.}~\bibnamefont{Lopez-Honorez}},
  \bibnamefont{and} \bibinfo{author}{\bibfnamefont{O.}~\bibnamefont{Mena}},
  \bibinfo{journal}{Phys.Rev.} \textbf{\bibinfo{volume}{D88}},
  \bibinfo{pages}{023531} (\bibinfo{year}{2013}).

\bibitem[{\citenamefont{Pettorino}(2013)}]{Pettorino:2013oxa}
\bibinfo{author}{\bibfnamefont{V.}~\bibnamefont{Pettorino}},
  \bibinfo{journal}{Phys.Rev.} \textbf{\bibinfo{volume}{D88}},
  \bibinfo{pages}{063519} (\bibinfo{year}{2013}).

\bibitem[{\citenamefont{Salvatelli et~al.}(2014)\citenamefont{Salvatelli, Said,
  Bruni, Melchiorri, and Wands}}]{Salvatelli:2014zta}
\bibinfo{author}{\bibfnamefont{V.}~\bibnamefont{Salvatelli}},
  \bibinfo{author}{\bibfnamefont{N.}~\bibnamefont{Said}},
  \bibinfo{author}{\bibfnamefont{M.}~\bibnamefont{Bruni}},
  \bibinfo{author}{\bibfnamefont{A.}~\bibnamefont{Melchiorri}},
  \bibnamefont{and} \bibinfo{author}{\bibfnamefont{D.}~\bibnamefont{Wands}},
  \bibinfo{journal}{Phys.Rev.Lett.} \textbf{\bibinfo{volume}{113}},
  \bibinfo{pages}{181301} (\bibinfo{year}{2014}).

\bibitem[{\citenamefont{Abdalla et~al.}(2014)\citenamefont{Abdalla, Ferreira,
  Quintin, and Wang}}]{Abdalla:2014cla}
\bibinfo{author}{\bibfnamefont{E.}~\bibnamefont{Abdalla}},
  \bibinfo{author}{\bibfnamefont{E.~G.~M.} \bibnamefont{Ferreira}},
  \bibinfo{author}{\bibfnamefont{J.}~\bibnamefont{Quintin}}, \bibnamefont{and}
  \bibinfo{author}{\bibfnamefont{B.}~\bibnamefont{Wang}}
  (\bibinfo{year}{2014}), \eprint{1412.2777}.

\bibitem[{\citenamefont{Avelino et~al.}(2006)\citenamefont{Avelino, Martins,
  Nunes, and Olive}}]{Avelino:2006gc}
\bibinfo{author}{\bibfnamefont{P.~P.} \bibnamefont{Avelino}},
  \bibinfo{author}{\bibfnamefont{C.~J. A.~P.} \bibnamefont{Martins}},
  \bibinfo{author}{\bibfnamefont{N.~J.} \bibnamefont{Nunes}}, \bibnamefont{and}
  \bibinfo{author}{\bibfnamefont{K.~A.} \bibnamefont{Olive}},
  \bibinfo{journal}{Phys.Rev.} \textbf{\bibinfo{volume}{D74}},
  \bibinfo{pages}{083508} (\bibinfo{year}{2006}).

\bibitem[{\citenamefont{Wetterich}(2007)}]{Wetterich:2007kr}
\bibinfo{author}{\bibfnamefont{C.}~\bibnamefont{Wetterich}},
  \bibinfo{journal}{Phys.Lett.} \textbf{\bibinfo{volume}{B655}},
  \bibinfo{pages}{201} (\bibinfo{year}{2007}).

\bibitem[{\citenamefont{Ayaita et~al.}(2012)\citenamefont{Ayaita, Weber, and
  Wetterich}}]{Ayaita:2011ay}
\bibinfo{author}{\bibfnamefont{Y.}~\bibnamefont{Ayaita}},
  \bibinfo{author}{\bibfnamefont{M.}~\bibnamefont{Weber}}, \bibnamefont{and}
  \bibinfo{author}{\bibfnamefont{C.}~\bibnamefont{Wetterich}},
  \bibinfo{journal}{Phys.Rev.} \textbf{\bibinfo{volume}{D85}},
  \bibinfo{pages}{123010} (\bibinfo{year}{2012}).

\bibitem[{\citenamefont{Faraoni et~al.}(2014)\citenamefont{Faraoni, Dent, and
  Saridakis}}]{Faraoni:2014vra}
\bibinfo{author}{\bibfnamefont{V.}~\bibnamefont{Faraoni}},
  \bibinfo{author}{\bibfnamefont{J.~B.} \bibnamefont{Dent}}, \bibnamefont{and}
  \bibinfo{author}{\bibfnamefont{E.~N.} \bibnamefont{Saridakis}},
  \bibinfo{journal}{Phys.Rev.} \textbf{\bibinfo{volume}{D90}},
  \bibinfo{pages}{063510} (\bibinfo{year}{2014}).

\end{thebibliography}

\end{document}